\documentclass[10pt, a4paper]{article}
\usepackage{verbatim}
\usepackage{color}
\definecolor{ao}{rgb}{0.0, 0.5, 0.0}
\usepackage[normalem]{ulem}
\usepackage{lipsum}
\usepackage{placeins}
\usepackage{bm}
\usepackage{textcase}
\usepackage[inkscapelatex=false]{svg}
\usepackage{arydshln}
\usepackage{nicefrac}
\usepackage{titlesec}
\usepackage{relsize}
\titleformat{\title}{\bfseries\large\centering}{}{0em}{}
\titlespacing*{\title}{0pt}{0pt}{12pt}

\usepackage[numbers,sort&compress]{natbib}
\usepackage{url}
\usepackage{hyperref}
\usepackage{amssymb}
\usepackage{fancyvrb}

\title{Synthesis and structural validation of close-to-stoichiometric NiTe$_2$ single crystals}
\author{T. Tirasutt  $^{1}$  , L. H. Tjeng  $^{1}$  and A. C. Komarek  $^{1,\Large{*}}$  \\
 $^{1}$: \small{Max Planck Institute for Chemical Physics of Solids, } \\
             \small{N{\"o}thnitzer Str. 40, 01187 Dresden, Germany} \\
 $^{\Large{*}}$: \small{Komarek@cpfs.mpg.de}
 }

\date{}

\begin{document}
\maketitle

\begin{abstract}
{We report the synthesis of close-to-stoichiometric NiTe$_{2}$ single crystals via chemical vapor transport together with precise structural characterization. A combined analysis using powder and single crystal X-ray diffraction (XRD) was performed to determine lattice parameters and structural details. High-resolution single crystal XRD reveals no evidence of interstitial nickel, confirming a close-to-stoichiometric composition. This work establishes a reliable benchmark for the crystal structure of NiTe$_{2}$, offering a reference for future studies on its physical properties.}
\end{abstract}

\section{Introduction}

Transition metal dichalcogenides (TMDCs), as a family of two-dimensional van der Waals materials, exhibit highly tunable electronic properties  \cite{Wang2012, Kuc2015, Manzeli2017}.
This tunability enables a broad range of potential applications, but also requires precise control and a thorough understanding of the material's intrinsic properties.
The electronic and physical characteristics of TMDCs depend not only on the choice of transition metal and chalcogen atoms, but also vary significantly with thickness, crystal structure, and stoichiometry.
For example, in Cr$_{\rm 1+x}$Te$_{2}$, the ferromagnetic Curie temperature ranges from 170~K to 350~K, depending on the degree of Cr intercalation \cite{Fujisawa2020, Zhang2025}.

\begin{figure}[!h]
    \centering
    \includegraphics[width=0.8\columnwidth]{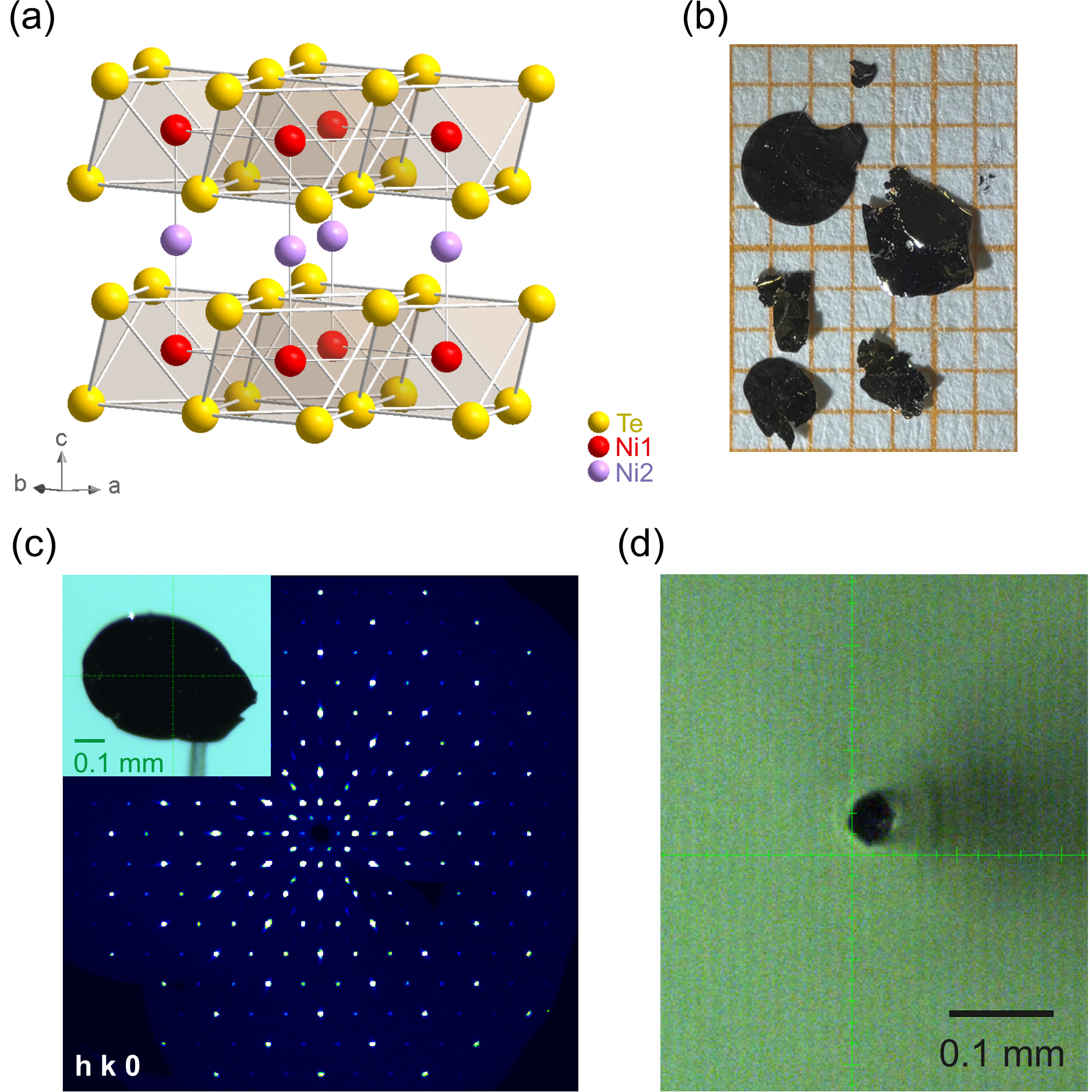}
    \caption{(a) The crystal structure of 1\textit{T}-NiTe$_{2}$  in the van der Waals layered phase, with the unit cell shown
(yellow: Te1, red: Ni1, violet: (empty) interstitial Ni2 site). (b) Photograph of as-grown, millimeter-sized NiTe$_{2}$ crystals on mm-paper. (c) Single crystal XRD measurements of a sufficiently large crystal (crystal~\#0) with an approximately elliptical, plate-like morphology (see inset), an average diameter of $\sim$0.5~mm, and a thickness of $\sim$20-25~$\mu$m confirm its single crystallinity. (d) Photograph of a smaller single crystal (crystal~\#1) used for accurate determination of the crystal structure and composition of NiTe$_{2}$ by single crystal XRD. }
    \label{figA}
\end{figure}

Here, we focus on the 3\textit{d} TMDC nickel ditelluride (NiTe$_{2}$), which has been reported as a semimetal exhibiting non-trivial topological surface states and is considered a candidate for a type-II Dirac semimetal, with the type-II Dirac point located near the Fermi level \cite{deLima2018, Xu2018, Ghosh2019, Mukherjee2020, Nurmamat2021, Bhatt2025}.
Fig.~\ref{figA}(a) illustrates the crystal structure of NiTe$_{2}$, which adopts a 1\textit{T}-type van der Waals layered structure (space group \textit{P}$\bar{3}$\textit{m}1, No. 164). Within each layer, Ni atoms are surrounded by Te atoms in an octahedral coordination environment with compressive trigonal distortion, consistent with \textit{D}$_{3d}$ point-group symmetry. Adjacent layers are stacked along the \textit{c}-axis and held together by weak van der Waals interactions.
A superconducting transition was observed in Te-deficient NiTe$_{\rm 2-x}$ (x~=~0.38~$\pm$~0.09) under hydrostatic pressure, with a critical temperature of \textit{T}$_{\rm C}$~=~7.5~K at 47.8~GPa \cite{Feng2021}.
Upon Ti intercalation at interstitial layer sites in NiTe$_{\rm 1.5}$, i.e. in Ti$_{0.1}$NiTe$_{1.5}$ \cite{deLima2018}, and Re doping in NiTe$_{\rm 2}$, i.e. in Ni$_{0.92}$Re$_{0.099}$Te$_2$ and Ni$_{0.87}$Re$_{0.13}$Te$_2$ \cite{Mandal2021}, the critical temperatures were observed to be approximately 4~K and 2~K, respectively.
More recently, in contrast to some theoretical predictions \cite{Zheng2020_theory} and electrical transport measurements~\cite{Esin2022,McFarlane2026}, intrinsic superconductivity was found in pristine stoichiometric samples below \textit{T}$_{\rm C}$~=~261~mK \cite{He2025}. NiTe$_{2}$ exhibits Pauli paramagnetic behavior, as evidenced by magnetic susceptibility measurements showing a small, temperature-independent response up to 300~K \cite{Mao2020, Zheng2020, He2025}. Overall, these observations underscore the importance of stoichiometry in the properties of Ni$_{1+\mathrm{x}}$Te$_{2}$ and suggest that deviations from ideal stoichiometry, such as the presence of excess Ni in interstitial sites within the layered crystal structure, can profoundly affect both superconducting and topological properties.

Despite numerous studies on this compound \cite{Mao2020}, most structural information remains limited to powder X-ray diffraction (XRD) or simple $\Theta$–$2\Theta$ scans on single crystal samples, which probe only the (0~0~l) reflection direction.
There are only a few reports on detailed single crystal XRD analyses of NiTe$_{2}$, all of which describe non-stoichiometric samples with excess 3\% -- 8\% interstitial nickel~\cite{Bensch1996, Zheng2020}. The interstitial Ni2 site is shown in Fig.~\ref{figA}(a).  
Doubts regarding the existence of stoichiometric NiTe$_{2}$ have also been raised \cite{Bensch1996}.

In this study, we present the synthesis of NiTe$_2$ single crystals via chemical vapor transport. Single crystal XRD analysis at high precision shows no occupation of interstitial sites by nickel nor sizable off-stoichiometry in the Ni:Te ratio, thereby confirming a close-to-stoichiometric composition.

\section{Materials and Methods}

\textbf{Chemical vapor transport (CVT)}\\
NiTe$_{2}$ single crystals were synthesized via CVT using iodine as the transporting agent. In an Ar-filled glovebox, a 1 g precursor pellet was prepared by thoroughly grinding Ni powder (Thermo Scientific, 99.996\%) and Te powder (Alfa Aesar, 99.99\%) in a 1:2 molar ratio. The pellet was then subjected to a pre-reaction in an evacuated quartz ampoule at 550$^{\circ}$C \cite{Komarek} for 3~days. After cooling to room temperature, the precursor was transferred into a new evacuated quartz ampoule (with about 17~cm length and 16~mm inner diameter), where approximately 100~mg of iodine was added to the ampoule with an inner volume of about 34~cm$^3$.
The ampoule was evacuated four times to $\sim$10$^{-1}$~mbar, with three intermediate backfills of Ar (99.999\% purity) to $\sim$1 bar.
Subsequently, CVT was performed in a two-zone furnace, with the hot zone maintained at 800$^{\circ}$C and the cold zone at 650$^{\circ}$C for 200~hours. The initial pressure of the transport additive was $\lesssim$~1~bar.
Before initiating the main transport from the hot zone to the cold zone, an initial back-transport step was carried out for
$\sim$1/2 d with the temperature gradient reversed. The initial heating to the maximum temperatures was completed within 3~h; after the back-transport step, the temperature gradient was reversed again within 2~h to initiate the main transport. After completion of the transport reaction, the ampoule was furnace-cooled to room temperature.
 \\

\textbf{Energy-dispersive X-ray (EDX) measurements }\\
EDX measurements were performed using a \textit{Bruker Quantax 400} system equipped with the \textit{Esprit} software package (version 2.3) and an \textit{XFlash 6|30} silicon-drift detector (SDD) mounted on a \textit{JSM-7800F} scanning electron microscope (\textit{JEOL}). Semi-quantitative, standardless analyses were carried out at an accelerating voltage of 20~kV. The chemical composition was calculated from the Ni K and Te L X-ray lines using the P/B-ZAF matrix-correction method.
A total of 20 data points were acquired in two measurements (10 points each), yielding a mean composition of
Ni$_{0.93(4)}$Te$_2$, where the value in parentheses denotes the standard deviation of all 20 measurements.
 \\

\textbf{X-ray diffraction (XRD) measurements}\\
Powder XRD measurements were carried out at room temperature using a \textit{Bruker D8 Discover} X-ray diffractometer equipped with Cu-K$_{\alpha 1}$ radiation. The sample was prepared by grinding intergrown single crystals into a fine powder prior to measurement.

Powder X-ray diffraction data were analysed by Rietveld refinement using \textit{FullProf/WinPLOTR}. Peak profiles were described by a pseudo-Voigt function with refined $U$, $V$, $W$ and $X$ width parameters and refined asymmetry parameters ($Y$ was fixed at zero).
The background was modelled using manually selected, interpolated background points, including a dense set of points between
2$\Theta$~$\sim$~22.7$^{\circ}$ and 28.2$^{\circ}$ to account for the broad contribution from the holder used for air-sensitive samples.
Preferred orientation was modelled along [0 0 1], consistent with the plate-like morphology of the layered NiTe$_2$ crystallites.
The refined parameters comprised the scale factor, lattice parameters, $z$(Te1), isotropic displacement parameters,
profile and asymmetry parameters, and preferred-orientation parameters. The goodness-of-fit is defined as $\chi^2=(\mathrm{R}_\mathrm{wp}/\mathrm{R}_\mathrm{exp})^2$,
and R$_\mathrm{Bragg}$ denotes the agreement between the observed and calculated integrated Bragg intensities. The refinement yielded
$\chi^2$~= 1.74 and R$_\mathrm{Bragg}$~= 2.65\%.

Single crystal X-ray diffraction measurements were performed at room temperature on a \emph{Bruker D8 VENTURE} single crystal X-ray diffractometer equipped with a bent graphite monochromator (Mo K$_{\alpha}$ radiation) and a \emph{Photon III} CMOS large-area detector. This compound is highly ductile, and single crystals deform readily under minimal mechanical stress.
An initial single crystal X-ray diffraction measurement was performed on a larger as-grown crystal (crystal~\#0) to confirm the single crystallinity throughout its volume, see Fig.~\ref{figA}(c).
The intensities of many reflections were so strong that the detector became saturated. Therefore, crystal~\#0 had to be re-measured using an attenuated incident beam.
The obtained data show 95\% (98\%) completeness up to 2$\Theta_\mathrm{max}$~= 144.2$^{\circ}$ (2$\Theta_\mathrm{full}$~= 132.6$^{\circ}$), with R$_\mathrm{int}$~=~4.87\%  and a redundancy of 22.4.
For the relatively large crystal (volume of about 0.0045~mm$^3$) of this Te-containing compound, very substantial absorption and extinction effects were observed. Absorption was corrected using the Gaussian integration method, yielding transmission factors of T$_\mathrm{min}$~= 0.057 and T$_\mathrm{max}$~= 0.558 (T$_\mathrm{min}$/T$_\mathrm{max}$~= 0.102); the Becker–Coppens type-1 Lorentzian extinction parameter was G$_\mathrm{iso}$~= 0.247(23). (An anisotropic extinction correction yielded only marginal improvement.) The crystal structure could be refined satisfactorily in the space group {\textit{P}$\bar{3}$\textit{m}1}, with GoF~=3.39, R=2.52\%, R$_{\mathrm{w}}$~= 7.53\%, and $z$(Te1)~= 0.251403(24), thereby further corroborating the single crystallinity of crystal~\#0. However, owing to systematic errors presumably introduced by the challenging absorption correction required for the relatively large crystal~\#0 (T$_\mathrm{min}$/T$_\mathrm{max}$~=~0.102), measurements of other, smaller specimens were required to resolve more subtle structural details, particularly for the refinement of site occupancies.

For more accurate single crystal X-ray diffraction measurements, small, thin, plate-like specimens were prepared from separate single crystals (crystals \#1, \#2 and \#3), whereas a different small, as-grown single crystal was measured directly (crystal~\#4). Crystals \#2 - \#4 were measured solely for comparison with crystal~\#1 and to assess the reproducibility of the refinement results across independently selected specimens from different single crystals. Owing to the appreciable X-ray absorption by Te, crystal~\#1 was selected as the reference specimen because it showed the smallest absorption variation, with T$_\mathrm{min}$/T$_\mathrm{max}$ = 0.703, compared with 0.567, 0.603 and 0.502 for crystals \#2, \#3 and \#4 respectively.
In addition, no significant extinction was detected for the extremely small crystal~\#1 ( $\sim$50~$\mu$m $\times$ $\sim$40~$\mu$m $\times$ $\lesssim$10~$\mu$m ), whereas the two larger crystals \#2 and \#4 showed appreciable extinction (see below).
Unless stated otherwise, the structural refinement results presented below always refer to crystal~\#1. (Crystals \#1 - \#4 are from the same batch.)
After integration of the intensities, a multi-scan absorption correction (\textit{SADABS 2016/2} \cite{sadabs}) was applied to the final data set, and the structure was refined using the crystallographic software \textit{Jana2006} \cite{Jana}.
Possible extinction was tested using an isotropic Becker–Coppens type-1 Lorentzian model. For crystal~\#1, the refined extinction parameter was indistinguishable from zero, G$_\mathrm{iso}$ = -0.000129(909), and inclusion of the extinction correction did not meaningfully change the refinement statistics (R/R$_\mathrm{w}$ = 2.45\%/5.51\% without, compared with 2.45\%/5.50\% with extinction correction). No extinction correction was therefore applied in the final refinement of the very thin crystal~\#1. (In contrast, e.g. the larger crystal~\#4  yielded already G$_\mathrm{iso}$ = 0.0646(29), i.e. a value significantly different from zero by approximately 23 $\sigma$, confirming appreciable extinction in this larger specimen.)
Structural refinement was performed by full-matrix least squares against F$_\mathrm{o}^2$, using all 330 independent reflections and weights
$w = (\sigma^2(\mathrm{I})+0.0004 \cdot\mathrm{I}^2)^{-1}$. The criterion $\mathrm{I} > 3\cdot\sigma(\mathrm{I})$ was used only to define the subset of 289 observed independent reflections for reporting R and R$_\mathrm{w}$; no reflections were excluded from the least-squares refinement. No restraints were applied. Standard uncertainties were derived from the least-squares variance–covariance matrix and represent formal uncertainties based on the adopted weighting scheme. In the subsequent occupancy refinements of Ni1 and/or Ni2, the anisotropic displacement parameters of Ni2 were constrained to the corresponding values of Ni1. The refinement results and additional measurement parameters are summarized in Tables~\ref{TxrdA}-\ref{TxrdC}.
The CIF, including the structure factors, and the \textit{checkCIF} alerts for crystal~\#1,
generated using the ‘Full validation of CIF and structure factors’ option, as well as the CIFs
and \textit{checkCIF} alerts for the other crystals, are provided in the Supplementary Material.

\section{Results and Discussion}

\begin{figure}[!t]
    \centering
    \includegraphics[width=1\columnwidth]{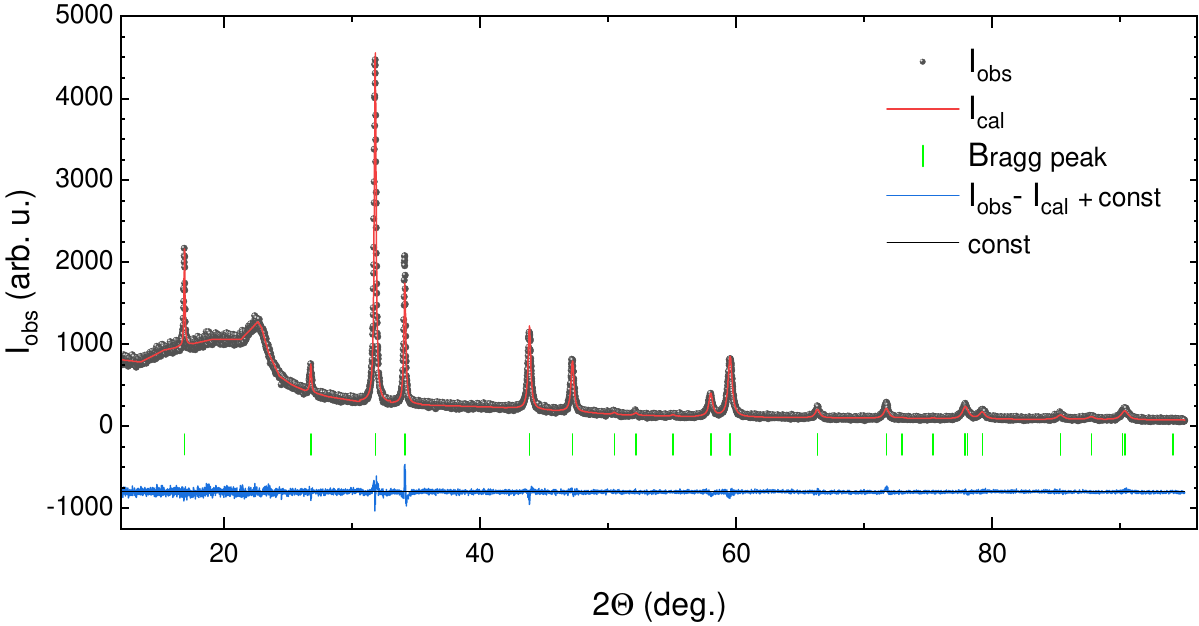}
    \caption{Powder X-ray diffraction pattern of NiTe$_{2}$ and the corresponding Rietveld refinement. Broad features below 2$\Theta$~$\approx$~26$^{\circ}$ arise from the holder used for air-sensitive samples. The refinement yielded $\chi^2$ = 1.74, R$_\mathrm{Bragg}$ = 2.65\%, and $z$(Te1)~=~0.2534(4).}
    \label{figB}
\end{figure}

\begin{figure}[!b]
    \centering
    \includegraphics[width=1\columnwidth]{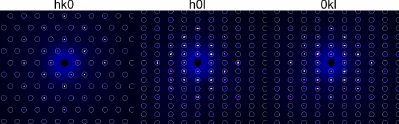}
    \caption{Precession images reconstructed from single crystal X-ray diffraction data of NiTe$_2$ (crystal~\#1), showing the ($hk0$), ($h0l$), and ($0kl$) planes of reciprocal space.}
    \label{figC}
\end{figure}

\begin{figure}[!t]
    \centering
    \includegraphics[width=0.64\columnwidth]{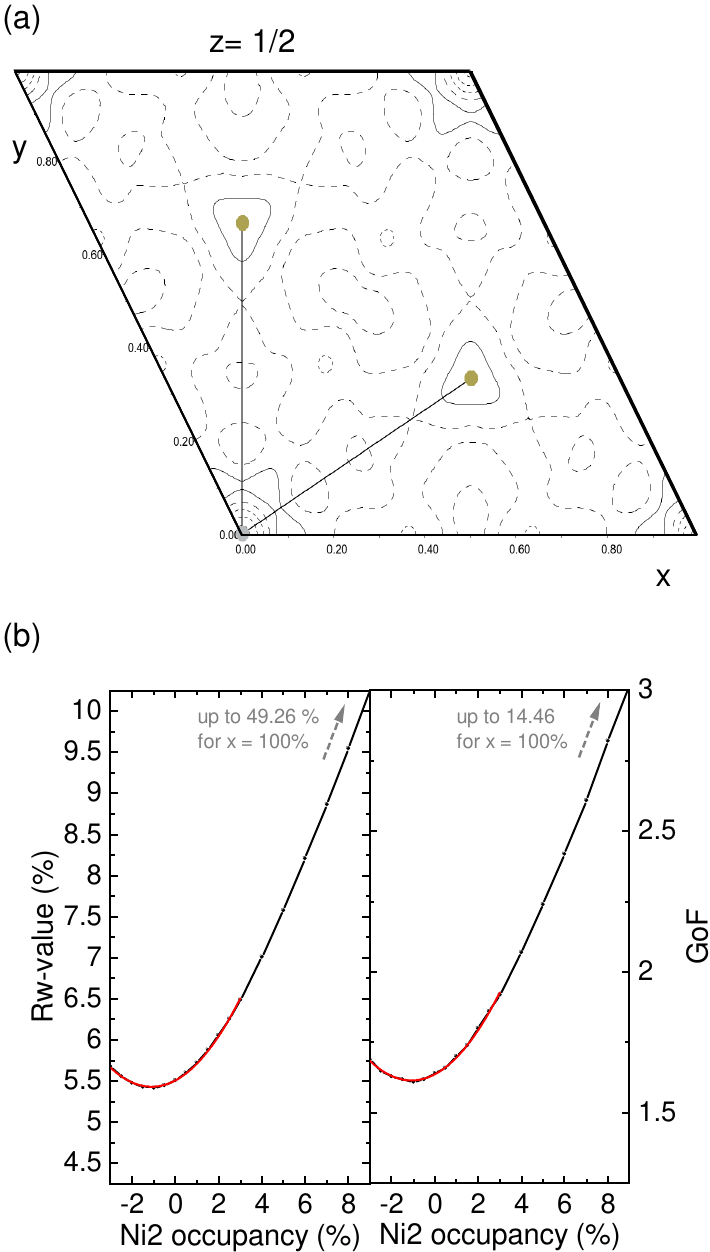}
    \caption{(a) Difference Fourier map showing $\Delta\rho(x,y,z)$ in the interstitial Ni layer ($z$~=~1/2) for NiTe$_2$ (crystal~\#1). The gray dot marks the position of the interstitial Ni2 site, while the other dots indicate the projected positions of the Te ions located at $z$(Te) and $1-z$(Te) outside this plane. Contour lines are drawn at intervals of $\pm$1~e$^-$/\AA$^3$; solid and short-dashed lines represent positive and negative contours, respectively, and the long-dashed line indicates the zero contour. (b) Weighted R-value and Goodness of fit plotted as a function of fixed Ni2 occupancy x, with all other parameters re-refined at each step (crystal~\#1).  The red lines are parabolic fits to the data in the range $-3\% \leq x \leq 3\%$.}
    \label{figD}
\end{figure}

Single crystals of NiTe$_{2}$ were synthesized via CVT (see the Methods section).
Approximately 20–30 shiny, plate-like crystals, each millimeter-sized and weighing on the order of a milligram, were obtained -- see the photograph in Fig.~\ref{figA}(b).
The single crystallinity of a sufficiently large crystal was confirmed by single crystal XRD measurements -- see Fig.~\ref{figA}(c).
The shiny appearance of the crystals persisted for more than one year in air, indicating no visible degradation or decomposition in air.
Energy dispersive X-ray spectroscopy (EDX) analysis yielded a composition of Ni$_{0.93(4)}$Te$_2$. Considering the limited absolute accuracy of EDX measurements performed without standards, these results should be interpreted with caution. Nevertheless, they are broadly consistent with the nominal 1:2 stoichiometry of the NiTe$_{2}$ crystals grown in this study.

\begin{table}[t]
  \begin{tabular}{llll}
   \hline
 occ. (Ni1)    & occ.(Ni2) & z(Te1)     &   R     / R$_w$  ($I > 3\sigma(I)$) \\
   1           & 0            & 0.25148(6) &  2.45\% / 5.51\% \\
   0.9897(57)   & 0            & 0.25149(6) &  2.45\% / 5.48\% \\
   1           & -0.0108(35)    & 0.25148(6) &  2.45\% / 5.42\% \\
   0.9908(57)    & -0.0104(35)    & 0.25148(6) &  2.44\% / 5.40\% \\
    \hline
 \end{tabular}
  \caption{Summary of most important structural refinement results for different refinements with or without refinement of the occupancy of Ni1 or Ni2 for the measurement of crystal~\#1. The anisotropic displacement parameters $U_\mathrm{i,j}$ of Ni2 were restricted to the corresponding values of the Ni1 atom. The first model is the one presented in Table~\ref{TxrdC}, as well as in the CIF and \textit{checkCIF} alerts.
  }\label{TxrdA}
\end{table}

\begin{table}[!hb]
  \begin{tabular}{lllll}
   \hline \hline
  crystal No.           &     \# 1        & \# 2             & \# 3            &  \# 4               \\
  \hline
 Reflections            & 9372            & 13853            & 1554            & 34358   \\
 Refl. (obs.)           & 7003            & 13574            & 1422            & 32062    \\
 Indep. Refl.           & 330             & 326              & 187             & 464   \\
 Ind. R. (obs.)         & 289             & 324              & 184             & 463   \\
 Redundancy             & 28.4            & 42.5             & 8.3             & 74.0      \\
 2$\Theta_\mathrm{max}$ & 105.1$^{\circ}$ &  103.5$^{\circ}$ & 80.1$^{\circ}$  & 125$^{\circ}$  \\
 completeness           & 97.6\%          & 98.2\%           & 98.4\%          & 100\%           \\
 R$_\mathrm{int}$ (obs) (\%)  & 4.83            & 2.91             & 3.45            & 4.47            \\
 T$_\mathrm{min.}$:T$_\mathrm{max}$    & 0.703 & 0.567   & 0.603              & 0.502  \\
  \hline
 occ. (Ni1)             &  0.9908(57)     &  0.9983(44)      & 0.9851(91)      & 1.0054(40)      \\
 occ.(Ni2)              & -0.0104(35)     &  -0.0108(25)     & 0.0003(56)      & -0.0159(23)    \\
 z(Te1)                 &  0.25148(6)     & 0.25152(2)       & 0.25161(9)      & 0.25148(2)       \\
 GoF                    & 1.61            & 1.89             & 2.19            & 1.81            \\
 R (\%)                 & 2.44            & 1.17             & 2.67            & 1.82            \\
 R$_\mathrm{w}$ (\%)    & 5.40            & 4.34             & 6.87            & 4.48          \\
 \hline
 occ. (Ni1)             &  1              &  1               & 1               & 1              \\
 occ.(Ni2)              &  0              &  0               &  0              &  0              \\
 z(Te1)                 &  0.25148(6)     & 0.25152(2)       & 0.25161(9)     & 0.25147(2)       \\
 GoF                    & 1.64             & 1.93             & 2.20            & 1.91             \\
 R (\%)                 & 2.45            & 1.19             & 2.66            & 1.87           \\
 R$_\mathrm{w}$ (\%)    & 5.51            & 4.47             & 6.92            & 4.73          \\
 \hline \hline
 \end{tabular}
  \caption{Comparison of key structural refinement results for several distinct single crystal specimens. The upper part of the table summarizes the measurement parameters. The central part lists the results obtained from the simultaneous refinement of the Ni1 and Ni2 site occupancies; the anisotropic displacement parameters, $U_{\mathrm{i,j}}$, of Ni2 were constrained to the corresponding values for Ni1. The lower part summarizes the results obtained from refinements for the stoichiometric composition NiTe$_2$.  }\label{TxrdB}
\end{table}

Fig.~\ref{figB} displays the powder X-ray diffraction data obtained in this study (see the Methods section for experimental details). Rietveld refinement of the diffraction pattern yielded $a = b$~= 3.85543(13)~\AA\ and $c$~= 5.26708(18)~\AA. The high angular resolution of the laboratory powder diffractometer allows the lattice parameters to be determined precisely from the peak positions. In contrast, the positional parameter $z$(Te) is less reliable because of the pronounced [0 0 1] preferred orientation of the plate-like crystallites. The refined preferred-orientation parameters differ substantially from the values expected in the absence of preferred orientation (Pref1 = 0.6449(983) $\neq$ 1), demonstrating strong texture, as expected for the plate-like morphology of the NiTe$_2$ crystallites. Consequently, the relative reflection intensities, and thus the powder-derived positional parameter, remain susceptible to systematic uncertainty despite the preferred-orientation correction. The lattice parameters are in good agreement with previously reported values~\cite{Xu2018, Mao2020, He2025}.

To determine the crystal structure in detail and address the question of Ni stoichiometry, single crystal XRD data were collected from a small, plate-like crystal of NiTe$_2$ (see the photograph in Fig.~\ref{figA}(d)). X-ray diffraction intensities in selected reciprocal-space planes are presented in Fig.~\ref{figC}. The experimental details and refinement results are summarized in Tab.~\ref{TxrdC}.

Single crystal X-ray diffraction is the method of choice for detecting small Ni occupancies at interstitial layer sites. In contrast to bulk compositional techniques such as EDX, single crystal X-ray diffraction yields site-specific electron density information (see e.g. Ref.~\cite{KomarekAC}), allowing for the resolution of local atomic environments, rather than providing only spatially averaged compositional data.
For example, in odd-order (0 0 2n+1) reflections, Ni atoms at the regular layer site ($z$~=~0) and at the interstitial site ($z$~=~1/2) scatter with opposite phases, leading to destructive interference rather than constructive addition.
We note that the scattering contribution from the two Te layers cancels exactly only for $z$(Te)~= 1/4. In practice, $z$(Te) is a freely refined structural parameter and may vary slightly with the interstitial Ni occupancy x through relaxation of the Te layers around the interstitial site. In the present refinements (Tabs.~\ref{TxrdA}--\ref{TxrdC}), however, $z$(Te)~$\approx$~1/4, and both $z$(Te) and x were refined simultaneously against the full diffraction dataset without sizable correlation.
Introducing a small interstitial occupancy x therefore directly suppresses the (0 0 2n+1) reflection intensities, following an
I~$\propto$~(1-x)$^2$ dependence. In this way, the (0 0 2n+1) reflections (and also many other reflections with l~=~2n+1) act as an intrinsic interferometric amplifier, translating a minute structural perturbation -- on the order of a few excess Ni atoms per hundred unit cells -- into a substantial relative change in reflection intensity, to the extent that these reflections vanish completely for x~=~1.
Consequently, these reflections serve as highly sensitive probes of interstitial Ni occupancy.
No equivalent mechanism exists in bulk-averaging techniques, which cannot decouple site-specific occupancies from the average composition and are therefore fundamentally limited in accuracy and are even prone to systematic errors.

\par Single crystal XRD has also been used in the literature to determine the presence of Ni atoms at the interstitial sites between the layers \cite{Bensch1996,Zheng2020}, and thus to establish the Ni excess in Ni$_{1+\mathrm{x}}$Te$_{2}$.
\par Although the EDX measurements suggested a slight Ni deficiency, their substantially larger uncertainty precluded a quantitative determination of the Ni:Te ratio. We therefore refined the occupancy of the regular Ni1 site. The refined value,
0.990(6), is close to unity, with full occupancy lying within approximately two standard uncertainties. Thus, the diffraction data provide no statistically significant evidence for Ni vacancies on the Ni1 site. Accordingly, the Ni1 occupancy was fixed at unity in the final refinement (Table~\ref{TxrdC}). For the present dataset, single crystal X-ray diffraction provides a more precise, site-specific assessment of the Ni1 occupancy, whereas EDX probes the average composition.

\begin{table}
  \begin{tabular}{l|l}
   \hline
Crystal   & NiTe$_{2}$  \\
Density & 7.6875 g/cm$^3$ \\
Temperature & room temperature \\
Wavelength & Mo K$_{\alpha}$ \\
Crystal system & trigonal \\
Space group & \textit{P $-3$ $m$ $1$} (164) \\
Unit cell dimensions\dag & $a$~=~3.85543(13) \AA \\
                         & $b$~=~3.85543(13) \AA \\
					     & $c$~=~5.26708(18) \AA \\
Volume\dag                   & 67.803(4) \AA$^3$ \\
Z & 1 \\
$F$(000)              & 132 \\
Crystal size  &   $\sim$50~$\mu$m $\times$ $\sim$40~$\mu$m $\times$ $\lesssim$10~$\mu$m  \\
$2$$\Theta_{max}$ & 105.1$^{\circ}$  \\
Index range   &  $h$: -8~$\rightarrow$~8 \\
              &  $k$: -8~$\rightarrow$~8 \\
			  &  $l$: -10~$\rightarrow$~11  \\
Reflections tot./indep. & 9372 /  330   \\
Obs. refl.  tot./indep. & 7003 / 289 \\
R$_\sigma$ (obs/all) & 5.43\% / 6.50\% \\
R$_\mathrm{int}$ (obs/all) &   4.83\%\ / 5.01\% \\
Redundancy & 28.4 \\
Completeness up to $2$$\Theta_{max}$ & 97.6\%  \\
Absorption correction & multi-scan \\
Min. / max. transmission &  0.703 / 1  \\
Refinement method & least squares on $F^2$ \\
Goodness of fit (obs/all)  &  1.72 / 1.64 \\
R / R$_w$ ($I > 3\sigma(I)$) &  2.45\% / 5.51\%  \\
R / R$_w$ (all) &  3.17\% / 5.58\%  \\
\hline
\textit{Positional parameters:} \\
\hdashline
Te1: \hspace{0.25cm}   $x$, $y$, $z$,                      &  0.3333, \hspace{0.83cm} 0.6667, \hspace{0.81cm} 0.25148(6)  \\
Ni1: \hspace{0.25cm}   $x$, $y$, $z$,                      &  0, \hspace{1.57cm} 0, \hspace{1.55cm} 0 \\
\hdashline
Te1: \hspace{0.25cm}   $occ.$,  $U_\mathrm{eq}$ (\AA$^2$) & 1,   \hspace{1.57cm}      0.01056(7) \\
Ni1: \hspace{0.25cm}   $occ.$,  $U_\mathrm{eq}$ (\AA$^2$) & 1,   \hspace{1.57cm}      0.00918(15) \\
\hdashline
Te1: \hspace{0.25cm}   $U_\mathrm{i,j}$ (\AA$^2$) & 0.00959(8), \hspace{0.25cm} 0.00959(8), \hspace{0.25cm}  0.01251(13), \\
      \hspace{0.25cm}                             & 0.00479(4), \hspace{0.26cm}  0, \hspace{1.57cm}  0 \\
Ni1: \hspace{0.25cm}   $U_\mathrm{i,j}$ (\AA$^2$) & 0.00818(16), \hspace{0.1cm}  0.00818(16), \hspace{0.09cm}  0.0112(3),\\
        \hspace{0.25cm}                           & 0.00409(8) ,  \hspace{0.15cm}  0, \hspace{1.59cm}  0 \\
\hline
 \end{tabular}
  \caption{Crystallographic and structural refinement data from a single crystal X-ray diffraction measurement of NiTe$_2$ (crystal~\#1). In addition to the scale factor, the refined parameters were $z$(Te1) and the anisotropic displacement parameters $U_\mathrm{1,1}$ and $U_\mathrm{3,3}$ for both Ni1 and Te1. \\ \dag : Lattice parameters taken from powder XRD.
  }\label{TxrdC}
\end{table}

To address the question of possible excess Ni at the interstitial layer sites, we first examined the
\textit{F}(obs)$-$\textit{F}(calc) difference-Fourier map at $z$~= 1/2, see Fig.~\ref{figD}(a). No positive residual electron density $\Delta\rho(x,y,z)$ is observed at
(0, 0, 1/2), a site previously identified as accommodating interstitial Ni ions \cite{Bensch1996}. Instead, the residual density at this position is slightly negative.
To quantify this residual scattering density, we subsequently refined the single crystal X-ray diffraction data by introducing an additional Ni2 atom at
(0, 0, 1/2), with its anisotropic displacement parameters constrained to those of Ni1. This refinement yielded a slightly negative formal occupancy of
-1.08(35)\%, consistent with the negative residual density and providing no evidence for interstitial Ni occupancy or excess Ni. Table~\ref{TxrdA} summarizes the key results of four refinements in which the occupancies of the Ni1 and/or Ni2 sites were either fixed or refined. The closely similar refined structural parameters and residual factors indicate that allowing for Ni1 deficiency or interstitial Ni2 occupancy does not significantly improve the fit, consistent with a Ni:Te ratio close to the ideal
1:2 stoichiometry within the sensitivity of the present single crystal X-ray diffraction data.

The refinement with freely refined Ni1 and Ni2 occupancies yields x = -1.04(35)\%\ (Tables~\ref{TxrdA} and \ref{TxrdB}), placing the physically meaningful value
x = 0 approximately 3$\sigma$ from the unconstrained minimum. To verify this formal result and explicitly account for the physical restriction
x~$\geq$~0, constrained profile refinements were performed with x fixed between -3\% and +9\%, while re-refining all remaining parameters. The anisotropic displacement parameters of Ni2 were constrained to the corresponding values of Ni1. The resulting R$_\mathrm{w}$(x) and GoF(x) profiles shown in Fig.~\ref{figD}(b) are parabolic near their minima at x $\sim$ -1.1\%. A Hamilton R-factor ratio test \cite{Hamilton} of the R$_\mathrm{w}$ profile (n = 289, p = 7, n-p = 282, and m = 1) confirms that the boundary value x = 0 remains compatible with the data at approximately the 3$\sigma$ level. Applying the physical restriction x $\geq$ 0 yields a formal one-sided upper limit of x $\lesssim$ +0.1\%. Thus, the data do not require a non-zero interstitial Ni2 occupancy and are consistent with NiTe$_2$ being close to stoichiometric, i.e. Ni$_{1+\mathrm{x}}$Te$_{2}$ with x $\sim$ 0, within the sensitivity of the present single crystal X-ray diffraction measurement.

Finally, refinements of three additional, independently selected single crystals (crystals \#2 - \#4) yielded closely comparable $z$(Te1) values and Ni1 and Ni2 occupancies statistically consistent with full Ni1-site occupation and zero interstitial Ni2 occupation (Table~\ref{TxrdB}). Despite the different data quality and resolution limits of the three measurements, the consistency of the results supports the conclusion that the investigated crystals are close to stoichiometric NiTe$_{2}$ within the sensitivity of the present single crystal X-ray diffraction measurements.
Therefore, the final structural refinements were performed with the Ni1 and Ni2 site occupancies fixed at 1 and 0, respectively (see Tables~\ref{TxrdB} and \ref{TxrdC} and the CIF provided in the Supplementary Material).

A possible explanation for the discrepancy between the interstitial Ni contents in the present work and in the study of Bensch \textit{et al.} \cite{Bensch1996}, which likewise employed CVT synthesis and single crystal XRD to assess the Ni content, may lie in the Ni-Te phase relations and the associated equilibrium decomposition pressure.
In the present CVT experiment, transport from the hot to the cold end may traverse the Te-rich boundary of the Ni$_{1+\mathrm{x}}$Te$_{2}$ homogeneity range, thereby favouring deposition of a Te-richer composition with a lower interstitial-Ni content.
On the other hand, in the explicitly described CVT synthesis of Bensch \textit{et al.} \cite{Bensch1996}, crystals were obtained in the hot zone, which, in this scenario, may correspond to transport across the Ni-rich boundary of the Ni$_{1+\mathrm{x}}$Te$_{2}$ homogeneity range and could favour a Ni-richer composition. The opposite transport directions in the present study and in the CVT synthesis of Bensch \textit{et al.} \cite{Bensch1996}, relative to the Te-rich and Ni-rich phase boundaries, may therefore contribute to the markedly lower interstitial Ni content of the present crystals.
In addition, the crystal growth temperature, precursor stoichiometry, and other transport conditions differ between the two studies. Hence, the difference in interstitial-Ni content of the obtained Ni$_{1+\mathrm{x}}$Te$_{2}$ single crystals in the two studies likely reflects the combined influence of several growth parameters rather than a single factor.

\section{Conclusions}
In conclusion, we present a study on the synthesis of NiTe$_{2}$ single crystals, accompanied by detailed structural characterization using X-ray diffraction (XRD). Powder XRD was used to determine the lattice parameters, while single crystal XRD provided high-resolution atomic-scale information.
The refined single crystal XRD data are consistent with a composition close to the ideal stoichiometry within experimental uncertainty.
The atomic parameters obtained are expected to be more reliable than those previously derived from powder XRD refinements.
This work establishes a reliable benchmark for the crystal structure of NiTe$_{2}$, serving as a reference for future experimental and theoretical investigations of this material.

\vspace{6pt}

\bibliographystyle{plainnat}
\bibliography{NiTe2}


\clearpage
\newpage

\textsc{\textbf{Supplementary Materials}} \\

\vspace{2cm}

\textsc{Content:}\\ \\
\begin{tabular}{cl}
1. & Summary of \textit{checkCIF} Alerts \\
2. & CIF of crystal \#1 \\
3. & CIF of crystal \#2 \\
4. & CIF of crystal \#3 \\
5. & CIF of crystal \#4 \\
6. & CIF of crystal \#0 \\

\end{tabular}

\vspace{3cm}

\textsc{1. Short summary of} \textit{checkCIF} \textsc{alerts:} \\ \\
\begin{tabular}{lcccc}
\hline
Crystal & A-level alerts & B-level alerts & C-level alerts & G-level alerts \\
\hline
Crystal \#1 & 0 & 0 & 0 & 8 \\
Crystal \#2 & 0 & 0 & 0 & 8 \\
Crystal \#3 & 0 & 0 & 3 & 8 \\
Crystal \#4 & 0 & 0 & 0 & 8 \\
Crystal \#0 \footnotemark & 2 & 1 & 3 & 9 \\
\hline
\end{tabular}

\footnotetext{This measurement was intended solely to demonstrate that this larger
as-grown crystal is also single crystalline. For specimens of this size containing the heavy
element Te, however, absorption effects are severe and difficult to correct adequately.}

\newpage

2. CIF of crystal \#1 : \\

\begin{Verbatim}
#=======================================================================
data_global
#=======================================================================

_audit_creation_method                   'Jana2006 Version : 04/12/2022'

#=======================================================================
data_I
#=======================================================================

# 5. CHEMICAL DATA

_chemical_formula_sum                    'Ni Te2'
_chemical_formula_weight                 313.9
_chemical_absolute_configuration         .

#=======================================================================
# 6. CRYSTAL DATA

_symmetry_cell_setting                   trigonal
_symmetry_space_group_name_H-M           'P -3 m 1'
_symmetry_space_group_name_Hall          '-P 3 2"'
_symmetry_Int_Tables_number              164

loop_
 _space_group_symop_id
 _space_group_symop_operation_xyz
 1   x,y,z
 2   -y,x-y,z
 3   -x+y,-x,z
 4   y,x,-z
 5   x-y,-y,-z
 6   -x,-x+y,-z
 7   -x,-y,-z
 8   y,-x+y,-z
 9   x-y,x,-z
 10  -y,-x,z
 11  -x+y,y,z
 12  x,x-y,z

_cell_length_a                           3.85540(10)
_cell_length_b                           3.85540(10)
_cell_length_c                           5.2671(2)
_cell_angle_alpha                        90
_cell_angle_beta                         90
_cell_angle_gamma                        120
_cell_volume                             67.802(4)
_cell_formula_units_Z                    1

_cell_measurement_reflns_used            1130
_cell_measurement_theta_min              9.3
_cell_measurement_theta_max              52.5
_cell_measurement_temperature            293

_exptl_crystal_density_diffrn            7.6875
_exptl_crystal_F_000                     132
_exptl_absorpt_coefficient_mu            27.861
_exptl_crystal_description               'plate-like shard'
_exptl_crystal_size_max                  0.05
_exptl_crystal_size_mid                  0.04
_exptl_crystal_size_min                  0.01
_exptl_crystal_colour                    'black'
_exptl_absorpt_correction_type           'multi-scan'
_exptl_absorpt_process_details           'SADABS 2016/2'
_exptl_absorpt_correction_T_min          0.7027
_exptl_absorpt_correction_T_max          1

#=======================================================================
# 7. EXPERIMENTAL DATA

_diffrn_ambient_temperature              293
_diffrn_source_voltage                   '50 kV'
_diffrn_source_current                   '30 mA'
_diffrn_radiation_type                   'Mo K\a'
_diffrn_radiation_source                 'X-ray tube'
_diffrn_radiation_wavelength             0.71075
_diffrn_radiation_monochromator          graphite
_diffrn_measurement_device               'four-circle diffractometer'
_diffrn_measurement_device_type          'Bruker D8 VENTURE'
_diffrn_detector                         'Photon III'
_diffrn_measurement_method               'phi and omega scans'

_diffrn_reflns_number                    9372
_diffrn_reflns_theta_min                 3.87
_diffrn_reflns_theta_max                 52.54
_diffrn_reflns_theta_full                52.41
_diffrn_measured_fraction_theta_max      0.98
_diffrn_measured_fraction_theta_full     0.98
_diffrn_reflns_av_R_equivalents          0.0501
_diffrn_reflns_av_unetI/netI             0.0191
_diffrn_reflns_limit_h_min               -8
_diffrn_reflns_limit_h_max               8
_diffrn_reflns_limit_k_min               -8
_diffrn_reflns_limit_k_max               8
_diffrn_reflns_limit_l_min               -10
_diffrn_reflns_limit_l_max               11
_diffrn_radiation_probe                  X-ray

#=======================================================================
# 8. REFINEMENT DATA

_refine_special_details
;
The Ni1 site occupancy was fixed at unity in the final structural model.
Anisotropic displacement parameters were refined for Ni1 and Te1.
The maximum residual electron-density peaks are located above and below
the Te position along the c axis.
;

_atom_sites_solution_primary             'charge flipping'
_atom_sites_solution_secondary           difmap
_atom_sites_solution_hydrogens           .

_reflns_number_total                     330
_reflns_number_gt                        289
_reflns_threshold_expression             'I>3\s(I)'

_refine_ls_structure_factor_coef         Fsqd
_refine_ls_R_factor_gt                   0.0245
_refine_ls_wR_factor_gt                  0.0551
_refine_ls_R_factor_all                  0.0317
_refine_ls_wR_factor_ref                 0.0558
_refine_ls_goodness_of_fit_ref           1.64
_refine_ls_goodness_of_fit_gt            1.72
_refine_ls_number_reflns                 330
_refine_ls_number_parameters             6
_refine_ls_number_restraints             0
_refine_ls_number_constraints            0
_refine_ls_weighting_scheme              sigma
_refine_ls_weighting_details             'w=1/(\s^2^(I)+0.0004I^2^)'
_refine_ls_shift/su_max                  0.0002
_refine_ls_shift/su_mean                 0.0000
_refine_diff_density_max                 3.71
_refine_diff_density_min                 -3.38
_refine_ls_extinction_method             'none'

loop_
 _atom_type_symbol
 _atom_type_scat_dispersion_real
 _atom_type_scat_dispersion_imag
 _atom_type_scat_source
 Ni   0.3393  1.1124
 'International Tables Vol C tables 4.2.6.8 and 6.1.1.1'
 Te  -0.5308  1.6751
 'International Tables Vol C tables 4.2.6.8 and 6.1.1.1'

_computing_data_collection               'APEX4'
_computing_cell_refinement               'SAINT'
_computing_data_reduction                'SAINT'
_computing_structure_solution            'SUPERFLIP'
_computing_structure_refinement          'Jana2006'

#=======================================================================
# 9. ATOMIC COORDINATES AND DISPLACEMENT PARAMETERS

loop_
 _atom_site_label
 _atom_site_type_symbol
 _atom_site_fract_x
 _atom_site_fract_y
 _atom_site_fract_z
 _atom_site_adp_type
 _atom_site_U_iso_or_equiv
 _atom_site_site_symmetry_multiplicity
 _atom_site_occupancy
 _atom_site_calc_flag
 _atom_site_refinement_flags
 _atom_site_disorder_assembly
 _atom_site_disorder_group
 Te1 Te 0.333333 0.666667 0.25148(6) Uani 0.01056(7) 2 1 d . . .
 Ni1 Ni 0 0 0 Uani 0.00918(15) 1 1 d . . .

loop_
 _atom_site_aniso_label
 _atom_site_aniso_type_symbol
 _atom_site_aniso_U_11
 _atom_site_aniso_U_22
 _atom_site_aniso_U_33
 _atom_site_aniso_U_12
 _atom_site_aniso_U_13
 _atom_site_aniso_U_23
 Te1 Te 0.00959(8) 0.00959(8) 0.01251(13) 0.00479(4) 0 0
 Ni1 Ni 0.00818(16) 0.00818(16) 0.0112(3) 0.00409(8) 0 0

#=======================================================================
# 10. MOLECULAR GEOMETRY

loop_
 _geom_bond_atom_site_label_1
 _geom_bond_atom_site_label_2
 _geom_bond_site_symmetry_1
 _geom_bond_site_symmetry_2
 _geom_bond_distance
 Te1 Ni1 . .     2.59022(16)
 Te1 Ni1 . 1_565 2.59022(16)
 Te1 Ni1 . 1_665 2.59022(16)

loop_
 _geom_angle_atom_site_label_1
 _geom_angle_atom_site_label_2
 _geom_angle_atom_site_label_3
 _geom_angle_site_symmetry_1
 _geom_angle_site_symmetry_2
 _geom_angle_site_symmetry_3
 _geom_angle
 Ni1 Te1 Ni1 .     . 1_565 96.185(8)
 Ni1 Te1 Ni1 .     . 1_665 96.184(8)
 Ni1 Te1 Ni1 1_565 . 1_665 96.185(8)
 Te1 Ni1 Te1 1_445 . 1_545 96.185(6)
 Te1 Ni1 Te1 1_445 . .     96.184(6)
 Te1 Ni1 Te1 1_445 . 4_445 83.816(6)
 Te1 Ni1 Te1 1_445 . 4_455 83.815(6)
 Te1 Ni1 Te1 1_445 . 4_555 180
 Te1 Ni1 Te1 1_545 . .     96.185(6)
 Te1 Ni1 Te1 1_545 . 4_445 83.815(6)
 Te1 Ni1 Te1 1_545 . 4_455 180
 Te1 Ni1 Te1 1_545 . 4_555 83.815(6)
 Te1 Ni1 Te1 .     . 4_445 180
 Te1 Ni1 Te1 .     . 4_455 83.815(6)
 Te1 Ni1 Te1 .     . 4_555 83.816(6)
 Te1 Ni1 Te1 4_445 . 4_455 96.185(6)
 Te1 Ni1 Te1 4_445 . 4_555 96.184(6)
 Te1 Ni1 Te1 4_455 . 4_555 96.185(6)
\end{Verbatim}
\newpage

3. CIF of crystal \#2 : \\

\begin{Verbatim}
#=======================================================================
data_global
#=======================================================================

_audit_creation_method                   'Jana2006 Version : 04/12/2022'

#=======================================================================
data_I
#=======================================================================

# 5. CHEMICAL DATA

_chemical_formula_sum                    'Ni Te2'
_chemical_formula_weight                 313.9
_chemical_absolute_configuration         .

#=======================================================================
# 6. CRYSTAL DATA

_symmetry_cell_setting                   trigonal
_symmetry_space_group_name_H-M           'P -3 m 1'
_symmetry_space_group_name_Hall          '-P 3 2"'
_symmetry_Int_Tables_number              164

loop_
 _space_group_symop_id
 _space_group_symop_operation_xyz
 1   x,y,z
 2   -y,x-y,z
 3   -x+y,-x,z
 4   y,x,-z
 5   x-y,-y,-z
 6   -x,-x+y,-z
 7   -x,-y,-z
 8   y,-x+y,-z
 9   x-y,x,-z
 10  -y,-x,z
 11  -x+y,y,z
 12  x,x-y,z

_cell_length_a                           3.85540(10)
_cell_length_b                           3.85540(10)
_cell_length_c                           5.2671(2)
_cell_angle_alpha                        90
_cell_angle_beta                         90
_cell_angle_gamma                        120
_cell_volume                             67.802(4)

loop_
 _twin_individual_id
 _twin_individual_mass_fraction_refined
 _twin_individual_twin_matrix_11
 _twin_individual_twin_matrix_12
 _twin_individual_twin_matrix_13
 _twin_individual_twin_matrix_21
 _twin_individual_twin_matrix_22
 _twin_individual_twin_matrix_23
 _twin_individual_twin_matrix_31
 _twin_individual_twin_matrix_32
 _twin_individual_twin_matrix_33
 1 1 1  0  0  0  1 0  0 0 1
 2 0 1 -1  0  1  0 0  0 0 1

_cell_formula_units_Z                    1

_cell_measurement_reflns_used            4968
_cell_measurement_theta_min              8.9
_cell_measurement_theta_max              51.7
_cell_measurement_temperature            293

_exptl_crystal_density_diffrn            7.6875
_exptl_crystal_F_000                     132
_exptl_absorpt_coefficient_mu            27.861
_exptl_crystal_description               'plate-like shard'
_exptl_crystal_size_max                  0.07
_exptl_crystal_size_mid                  0.05
_exptl_crystal_size_min                  0.013
_exptl_crystal_colour                    'black'
_exptl_absorpt_correction_type           'multi-scan'
_exptl_absorpt_process_details           'SADABS 2016/2'
_exptl_absorpt_correction_T_min          0.5673
_exptl_absorpt_correction_T_max          1

#=======================================================================
# 7. EXPERIMENTAL DATA

_diffrn_ambient_temperature              293
_diffrn_radiation_type                   'Mo K\a'
_diffrn_radiation_source                 'X-ray tube'
_diffrn_radiation_wavelength             0.71075
_diffrn_radiation_monochromator          graphite
_diffrn_measurement_device               'four-circle diffractometer'
_diffrn_measurement_device_type          'Bruker D8 VENTURE'
_diffrn_detector                         'Photon III'
_diffrn_measurement_method               'phi and omega scans'

_diffrn_reflns_number                    13853
_diffrn_reflns_theta_min                 3.87
_diffrn_reflns_theta_max                 51.74
_diffrn_reflns_theta_full                51.74
_diffrn_measured_fraction_theta_max      0.98
_diffrn_measured_fraction_theta_full     0.98
_diffrn_reflns_av_R_equivalents          0.0291
_diffrn_reflns_av_unetI/netI             0.0086
_diffrn_reflns_limit_h_min               -8
_diffrn_reflns_limit_h_max               8
_diffrn_reflns_limit_k_min               -8
_diffrn_reflns_limit_k_max               8
_diffrn_reflns_limit_l_min               -11
_diffrn_reflns_limit_l_max               11
_diffrn_radiation_probe                  X-ray

#=======================================================================
# 8. REFINEMENT DATA

_refine_special_details
;
The Ni1 site occupancy was fixed at unity in the final structural model.
Anisotropic displacement parameters were refined for Ni1 and Te1.
The maximum residual electron-density peaks are located above and below
the Te position along the c axis.
;

_atom_sites_solution_primary             'charge flipping'
_atom_sites_solution_secondary           difmap
_atom_sites_solution_hydrogens           .

_reflns_number_total                     326
_reflns_number_gt                        324
_reflns_threshold_expression             'I>3\s(I)'

_refine_ls_structure_factor_coef         Fsqd
_refine_ls_R_factor_gt                   0.0119
_refine_ls_wR_factor_gt                  0.0447
_refine_ls_R_factor_all                  0.0121
_refine_ls_wR_factor_ref                 0.0447
_refine_ls_goodness_of_fit_ref           1.93
_refine_ls_goodness_of_fit_gt            1.94
_refine_ls_number_reflns                 326
_refine_ls_number_parameters             7
_refine_ls_number_restraints             0
_refine_ls_number_constraints            0
_refine_ls_weighting_scheme              sigma
_refine_ls_weighting_details             'w=1/(\s^2^(I)+0.0004I^2^)'
_refine_ls_shift/su_max                  0.0005
_refine_ls_shift/su_mean                 0.0001
_refine_diff_density_max                 2.50
_refine_diff_density_min                 -3.70
_refine_ls_extinction_method
 'B-C type 1 Lorentzian isotropic (Becker & Coppens, 1974)'
_refine_ls_extinction_coef               190(30)

loop_
 _atom_type_symbol
 _atom_type_scat_dispersion_real
 _atom_type_scat_dispersion_imag
 _atom_type_scat_source
 Ni   0.3393  1.1124
 'International Tables Vol C tables 4.2.6.8 and 6.1.1.1'
 Te  -0.5308  1.6751
 'International Tables Vol C tables 4.2.6.8 and 6.1.1.1'

_computing_data_collection               'APEX4'
_computing_cell_refinement               'SAINT'
_computing_data_reduction                'SAINT'
_computing_structure_solution            'SUPERFLIP'
_computing_structure_refinement          'Jana2006'

#=======================================================================
# 9. ATOMIC COORDINATES AND DISPLACEMENT PARAMETERS

loop_
 _atom_site_label
 _atom_site_type_symbol
 _atom_site_fract_x
 _atom_site_fract_y
 _atom_site_fract_z
 _atom_site_adp_type
 _atom_site_U_iso_or_equiv
 _atom_site_site_symmetry_multiplicity
 _atom_site_occupancy
 _atom_site_calc_flag
 _atom_site_refinement_flags
 _atom_site_disorder_assembly
 _atom_site_disorder_group
 Te1 Te 0.333333 0.666667 0.25152(2) Uani 0.00943(5) 2 1 d . . .
 Ni1 Ni 0 0 0 Uani 0.00828(8) 1 1 d . . .

loop_
 _atom_site_aniso_label
 _atom_site_aniso_type_symbol
 _atom_site_aniso_U_11
 _atom_site_aniso_U_22
 _atom_site_aniso_U_33
 _atom_site_aniso_U_12
 _atom_site_aniso_U_13
 _atom_site_aniso_U_23
 Te1 Te 0.00882(6) 0.00882(6) 0.01065(8) 0.00441(3) 0 0
 Ni1 Ni 0.00733(9) 0.00733(9) 0.01018(13) 0.00366(4) 0 0

#=======================================================================
# 10. MOLECULAR GEOMETRY

loop_
 _geom_bond_atom_site_label_1
 _geom_bond_atom_site_label_2
 _geom_bond_site_symmetry_1
 _geom_bond_site_symmetry_2
 _geom_bond_distance
 Te1 Ni1 . .     2.59031(6)
 Te1 Ni1 . 1_565 2.59031(6)
 Te1 Ni1 . 1_665 2.59031(6)

loop_
 _geom_angle_atom_site_label_1
 _geom_angle_atom_site_label_2
 _geom_angle_atom_site_label_3
 _geom_angle_site_symmetry_1
 _geom_angle_site_symmetry_2
 _geom_angle_site_symmetry_3
 _geom_angle
 Ni1 Te1 Ni1 .     . 1_565 96.180(3)
 Ni1 Te1 Ni1 .     . 1_665 96.180(3)
 Ni1 Te1 Ni1 1_565 . 1_665 96.180(3)
 Te1 Ni1 Te1 1_445 . 1_545 96.180(2)
 Te1 Ni1 Te1 1_445 . .     96.180(2)
 Te1 Ni1 Te1 1_445 . 4_445 83.820(2)
 Te1 Ni1 Te1 1_445 . 4_455 83.820(2)
 Te1 Ni1 Te1 1_445 . 4_555 180
 Te1 Ni1 Te1 1_545 . .     96.180(2)
 Te1 Ni1 Te1 1_545 . 4_445 83.820(2)
 Te1 Ni1 Te1 1_545 . 4_455 180
 Te1 Ni1 Te1 1_545 . 4_555 83.820(2)
 Te1 Ni1 Te1 .     . 4_445 180
 Te1 Ni1 Te1 .     . 4_455 83.820(2)
 Te1 Ni1 Te1 .     . 4_555 83.820(2)
 Te1 Ni1 Te1 4_445 . 4_455 96.180(2)
 Te1 Ni1 Te1 4_445 . 4_555 96.180(2)
 Te1 Ni1 Te1 4_455 . 4_555 96.180(2)



\end{Verbatim}
\newpage

4. CIF of crystal \#3 : \\

\begin{Verbatim}
#=======================================================================
data_global
#=======================================================================

_audit_creation_method                   'Jana2006 Version : 04/12/2022'

#=======================================================================
data_I
#=======================================================================

# 5. CHEMICAL DATA

_chemical_formula_sum                    'Ni Te2'
_chemical_formula_weight                 313.9
_chemical_absolute_configuration         .

#=======================================================================
# 6. CRYSTAL DATA

_symmetry_cell_setting                   trigonal
_symmetry_space_group_name_H-M           'P -3 m 1'
_symmetry_space_group_name_Hall          '-P 3 2"'
_symmetry_Int_Tables_number              164

loop_
 _space_group_symop_id
 _space_group_symop_operation_xyz
 1   x,y,z
 2   -y,x-y,z
 3   -x+y,-x,z
 4   y,x,-z
 5   x-y,-y,-z
 6   -x,-x+y,-z
 7   -x,-y,-z
 8   y,-x+y,-z
 9   x-y,x,-z
 10  -y,-x,z
 11  -x+y,y,z
 12  x,x-y,z

_cell_length_a                           3.85540(10)
_cell_length_b                           3.85540(10)
_cell_length_c                           5.2671(2)
_cell_angle_alpha                        90
_cell_angle_beta                         90
_cell_angle_gamma                        120
_cell_volume                             67.802(4)

loop_
 _twin_individual_id
 _twin_individual_mass_fraction_refined
 _twin_individual_twin_matrix_11
 _twin_individual_twin_matrix_12
 _twin_individual_twin_matrix_13
 _twin_individual_twin_matrix_21
 _twin_individual_twin_matrix_22
 _twin_individual_twin_matrix_23
 _twin_individual_twin_matrix_31
 _twin_individual_twin_matrix_32
 _twin_individual_twin_matrix_33
 1 1 1  0  0  0  1 0  0 0 1
 2 0 1 -1  0  1  0 0  0 0 1

_cell_formula_units_Z                    1

_cell_measurement_reflns_used            3461
_cell_measurement_theta_min              9.1
_cell_measurement_theta_max              40.0
_cell_measurement_temperature            293

_exptl_crystal_density_diffrn            7.6875
_exptl_crystal_F_000                     132
_exptl_absorpt_coefficient_mu            27.861
_exptl_crystal_description               'plate-like shard'
_exptl_crystal_size_max                  0.13
_exptl_crystal_size_mid                  0.05
_exptl_crystal_size_min                  0.01
_exptl_crystal_colour                    'black'
_exptl_absorpt_correction_type           'multi-scan'
_exptl_absorpt_process_details           'SADABS 2016/2'
_exptl_absorpt_correction_T_min          0.6028
_exptl_absorpt_correction_T_max          1

#=======================================================================
# 7. EXPERIMENTAL DATA

_diffrn_ambient_temperature              293
_diffrn_radiation_type                   'Mo K\a'
_diffrn_radiation_source                 'X-ray tube'
_diffrn_radiation_wavelength             0.71075
_diffrn_radiation_monochromator          graphite
_diffrn_measurement_device               'four-circle diffractometer'
_diffrn_measurement_device_type          'Bruker D8 VENTURE'
_diffrn_detector                         'Photon III'
_diffrn_measurement_method               'phi and omega scans'

_diffrn_reflns_number                    1554
_diffrn_reflns_theta_min                 3.87
_diffrn_reflns_theta_max                 40.05
_diffrn_reflns_theta_full                40.05
_diffrn_measured_fraction_theta_max      0.98
_diffrn_measured_fraction_theta_full     0.98
_diffrn_reflns_av_R_equivalents          0.0346
_diffrn_reflns_av_unetI/netI             0.0193
_diffrn_reflns_limit_h_min               -6
_diffrn_reflns_limit_h_max               6
_diffrn_reflns_limit_k_min               -6
_diffrn_reflns_limit_k_max               6
_diffrn_reflns_limit_l_min               -9
_diffrn_reflns_limit_l_max               9
_diffrn_radiation_probe                  X-ray

#=======================================================================
# 8. REFINEMENT DATA

_refine_special_details
;
The Ni1 site occupancy was fixed at unity in the final structural model.
Anisotropic displacement parameters were refined for Ni1 and Te1.
The maximum residual electron-density peaks are located above and below
the Te position along the c axis.
;

_atom_sites_solution_primary             'charge flipping'
_atom_sites_solution_secondary           difmap
_atom_sites_solution_hydrogens           .

_reflns_number_total                     187
_reflns_number_gt                        184
_reflns_threshold_expression             'I>3\s(I)'

_refine_ls_structure_factor_coef         Fsqd
_refine_ls_R_factor_gt                   0.0266
_refine_ls_wR_factor_gt                  0.0692
_refine_ls_R_factor_all                  0.0273
_refine_ls_wR_factor_ref                 0.0694
_refine_ls_goodness_of_fit_ref           2.20
_refine_ls_goodness_of_fit_gt            2.21
_refine_ls_number_reflns                 187
_refine_ls_number_parameters             6
_refine_ls_number_restraints             0
_refine_ls_number_constraints            0
_refine_ls_weighting_scheme              sigma
_refine_ls_weighting_details             'w=1/(\s^2^(I)+0.0004I^2^)'
_refine_ls_shift/su_max                  0.0000
_refine_ls_shift/su_mean                 0.0000
_refine_diff_density_max                 4.07
_refine_diff_density_min                 -2.88
_refine_ls_extinction_method             'none'

loop_
 _atom_type_symbol
 _atom_type_scat_dispersion_real
 _atom_type_scat_dispersion_imag
 _atom_type_scat_source
 Ni   0.3393  1.1124
 'International Tables Vol C tables 4.2.6.8 and 6.1.1.1'
 Te  -0.5308  1.6751
 'International Tables Vol C tables 4.2.6.8 and 6.1.1.1'

_computing_data_collection               'APEX4'
_computing_cell_refinement               'SAINT'
_computing_data_reduction                'SAINT'
_computing_structure_solution            'SUPERFLIP'
_computing_structure_refinement          'Jana2006'

#=======================================================================
# 9. ATOMIC COORDINATES AND DISPLACEMENT PARAMETERS

loop_
 _atom_site_label
 _atom_site_type_symbol
 _atom_site_fract_x
 _atom_site_fract_y
 _atom_site_fract_z
 _atom_site_adp_type
 _atom_site_U_iso_or_equiv
 _atom_site_site_symmetry_multiplicity
 _atom_site_occupancy
 _atom_site_calc_flag
 _atom_site_refinement_flags
 _atom_site_disorder_assembly
 _atom_site_disorder_group
 Te1 Te 0.333333 0.666667 0.25161(9) Uani 0.00864(14) 2 1 d . . .
 Ni1 Ni 0 0 0 Uani 0.0074(3) 1 1 d . . .

loop_
 _atom_site_aniso_label
 _atom_site_aniso_type_symbol
 _atom_site_aniso_U_11
 _atom_site_aniso_U_22
 _atom_site_aniso_U_33
 _atom_site_aniso_U_12
 _atom_site_aniso_U_13
 _atom_site_aniso_U_23
 Te1 Te 0.00725(17) 0.00725(17) 0.0114(2) 0.00363(9) 0 0
 Ni1 Ni 0.0058(3) 0.0058(3) 0.0106(5) 0.00289(17) 0 0

#=======================================================================
# 10. MOLECULAR GEOMETRY

loop_
 _geom_bond_atom_site_label_1
 _geom_bond_atom_site_label_2
 _geom_bond_site_symmetry_1
 _geom_bond_site_symmetry_2
 _geom_bond_distance
 Te1 Ni1 . .     2.5906(2)
 Te1 Ni1 . 1_565 2.5906(2)
 Te1 Ni1 . 1_665 2.5906(2)

loop_
 _geom_angle_atom_site_label_1
 _geom_angle_atom_site_label_2
 _geom_angle_atom_site_label_3
 _geom_angle_site_symmetry_1
 _geom_angle_site_symmetry_2
 _geom_angle_site_symmetry_3
 _geom_angle
 Ni1 Te1 Ni1 .     . 1_565 96.168(12)
 Ni1 Te1 Ni1 .     . 1_665 96.167(12)
 Ni1 Te1 Ni1 1_565 . 1_665 96.168(12)
 Te1 Ni1 Te1 1_445 . 1_545 96.168(8)
 Te1 Ni1 Te1 1_445 . .     96.167(8)
 Te1 Ni1 Te1 1_445 . 4_445 83.833(8)
 Te1 Ni1 Te1 1_445 . 4_455 83.832(8)
 Te1 Ni1 Te1 1_445 . 4_555 180
 Te1 Ni1 Te1 1_545 . .     96.168(8)
 Te1 Ni1 Te1 1_545 . 4_445 83.832(8)
 Te1 Ni1 Te1 1_545 . 4_455 180
 Te1 Ni1 Te1 1_545 . 4_555 83.832(8)
 Te1 Ni1 Te1 .     . 4_445 180
 Te1 Ni1 Te1 .     . 4_455 83.832(8)
 Te1 Ni1 Te1 .     . 4_555 83.833(8)
 Te1 Ni1 Te1 4_445 . 4_455 96.168(8)
 Te1 Ni1 Te1 4_445 . 4_555 96.167(8)
 Te1 Ni1 Te1 4_455 . 4_555 96.168(8)


# start Validation Reply Form
_vrf_DIFMX02_I
;
PROBLEM: The maximum difference density is > 0.1*ZMAX*0.75
RESPONSE: The maximum residual-density peak (+4.07 eA-3) is located along
          the c direction from the Te1 site. Its value is only marginally
          above the checkCIF threshold for this Te-containing compound
          (0.1*ZMAX*0.75 = 3.90 eA-3); the minimum residual density is
          -2.88 eA-3. Crystal #3 was affected by slight
          bending/deformation during preparation, owing to the high ductility
          of the material. Consequently, higher-angle reflections became weak
          and broadened and were excluded from the refinement. The residual
          density is attributed to the limited high-angle data quality of the
          slightly deformed crystal.
;
_vrf_GOODF01_I
;
PROBLEM: The least squares goodness of fit parameter lies
RESPONSE: The moderately elevated goodness-of-fit value is attributed to the
          limited diffraction quality of crystal #3. Owing to the high
          ductility of the material, slight bending/deformation of the small
          crystal during preparation could not be avoided. Higher-angle
          reflections became weak and broadened; therefore, only data up to
          2Theta_max = 80.1 degrees were included in the refinement.
          The difference-Fourier extrema remain moderate (+4.07 and
          -2.88 eA-3), and the refinement converged satisfactorily.
;
_vrf_PLAT097_I
;
PROBLEM: Large Reported Max. (Positive) Residual Density 4.07 eA-3
RESPONSE: See the response to DIFMX02_ALERT_1_C. The reported positive
          residual density corresponds to the same difference-Fourier peak.
          At +4.07 eA-3, it is only marginally above the applicable
          checkCIF threshold (0.1*ZMAX*0.75 = 3.90 eA-3) and is attributed
          to the limited high-angle data quality of the slightly deformed
          crystal, owing to the high ductility of the material.
;
# end Validation Reply Form

\end{Verbatim}
\newpage

5. CIF of crystal \#4 : \\

\begin{Verbatim}
#=======================================================================
data_global
#=======================================================================

_audit_creation_method                   'Jana2006 Version : 04/12/2022'

#=======================================================================
data_I
#=======================================================================

# 5. CHEMICAL DATA

_chemical_formula_sum                    'Ni Te2'
_chemical_formula_weight                 313.9
_chemical_absolute_configuration         .

#=======================================================================
# 6. CRYSTAL DATA

_symmetry_cell_setting                   trigonal
_symmetry_space_group_name_H-M           'P -3 m 1'
_symmetry_space_group_name_Hall          '-P 3 2"'
_symmetry_Int_Tables_number              164

loop_
 _space_group_symop_id
 _space_group_symop_operation_xyz
 1   x,y,z
 2   -y,x-y,z
 3   -x+y,-x,z
 4   y,x,-z
 5   x-y,-y,-z
 6   -x,-x+y,-z
 7   -x,-y,-z
 8   y,-x+y,-z
 9   x-y,x,-z
 10  -y,-x,z
 11  -x+y,y,z
 12  x,x-y,z

_cell_length_a                           3.85540(10)
_cell_length_b                           3.85540(10)
_cell_length_c                           5.2671(2)
_cell_angle_alpha                        90
_cell_angle_beta                         90
_cell_angle_gamma                        120
_cell_volume                             67.802(4)

loop_
 _twin_individual_id
 _twin_individual_mass_fraction_refined
 _twin_individual_twin_matrix_11
 _twin_individual_twin_matrix_12
 _twin_individual_twin_matrix_13
 _twin_individual_twin_matrix_21
 _twin_individual_twin_matrix_22
 _twin_individual_twin_matrix_23
 _twin_individual_twin_matrix_31
 _twin_individual_twin_matrix_32
 _twin_individual_twin_matrix_33
 1 1 1  0  0  0  1 0  0 0 1
 2 0 1 -1  0  1  0 0  0 0 1
 3 0 0 -1  0  1 -1 0  0 0 1

_cell_formula_units_Z                    1

_cell_measurement_reflns_used            761
_cell_measurement_theta_min              9.3
_cell_measurement_theta_max              62.5
_cell_measurement_temperature            293

_exptl_crystal_density_diffrn            7.6875
_exptl_crystal_F_000                     132
_exptl_absorpt_coefficient_mu            27.861
_exptl_crystal_description               'flat cylindrical'
_exptl_crystal_size_max                  0.1
_exptl_crystal_size_mid                  0.1
_exptl_crystal_size_min                  0.015
_exptl_crystal_size_rad                  0.038
_exptl_crystal_colour                    'black'
_exptl_absorpt_correction_type           'multi-scan'
_exptl_absorpt_process_details           'SADABS 2016/2'
_exptl_absorpt_correction_T_min          0.5018
_exptl_absorpt_correction_T_max          1

#=======================================================================
# 7. EXPERIMENTAL DATA

_diffrn_ambient_temperature              293
_diffrn_radiation_type                   'Mo K\a'
_diffrn_radiation_source                 'X-ray tube'
_diffrn_radiation_wavelength             0.71075
_diffrn_radiation_monochromator          graphite
_diffrn_measurement_device               'four-circle diffractometer'
_diffrn_measurement_device_type          'Bruker D8 VENTURE'
_diffrn_detector                         'Photon III'
_diffrn_measurement_method               'phi and omega scans'

_diffrn_reflns_number                    34358
_diffrn_reflns_theta_min                 3.87
_diffrn_reflns_theta_max                 62.48
_diffrn_reflns_theta_full                62.48
_diffrn_measured_fraction_theta_max      1
_diffrn_measured_fraction_theta_full     1
_diffrn_reflns_av_R_equivalents          0.0447
_diffrn_reflns_av_unetI/netI             0.0079
_diffrn_reflns_limit_h_min               -9
_diffrn_reflns_limit_h_max               9
_diffrn_reflns_limit_k_min               -9
_diffrn_reflns_limit_k_max               9
_diffrn_reflns_limit_l_min               -13
_diffrn_reflns_limit_l_max               13
_diffrn_radiation_probe                  X-ray

#=======================================================================
# 8. REFINEMENT DATA

_refine_special_details
;
The Ni1 site occupancy was fixed at unity in the final structural model.
Anisotropic displacement parameters were refined for Ni1 and Te1.
The maximum residual electron-density peaks are located above and below
the Te position along the c axis.
;

_atom_sites_solution_primary             'charge flipping'
_atom_sites_solution_secondary           difmap
_atom_sites_solution_hydrogens           .

_reflns_number_total                     464
_reflns_number_gt                        463
_reflns_threshold_expression             'I>3\s(I)'

_refine_ls_structure_factor_coef         Fsqd
_refine_ls_R_factor_gt                   0.0187
_refine_ls_wR_factor_gt                  0.0473
_refine_ls_R_factor_all                  0.0188
_refine_ls_wR_factor_ref                 0.0473
_refine_ls_goodness_of_fit_ref           1.91
_refine_ls_goodness_of_fit_gt            1.91
_refine_ls_number_reflns                 464
_refine_ls_number_parameters             7
_refine_ls_number_restraints             0
_refine_ls_number_constraints            0
_refine_ls_weighting_scheme              sigma
_refine_ls_weighting_details             'w=1/(\s^2^(I)+0.0004I^2^)'
_refine_ls_shift/su_max                  0.0003
_refine_ls_shift/su_mean                 0.0000
_refine_diff_density_max                 3.19
_refine_diff_density_min                 -2.94
_refine_ls_extinction_method
 'B-C type 1 Lorentzian isotropic (Becker & Coppens, 1974)'
_refine_ls_extinction_coef               650(30)

loop_
 _atom_type_symbol
 _atom_type_scat_dispersion_real
 _atom_type_scat_dispersion_imag
 _atom_type_scat_source
 Ni   0.3393  1.1124
 'International Tables Vol C tables 4.2.6.8 and 6.1.1.1'
 Te  -0.5308  1.6751
 'International Tables Vol C tables 4.2.6.8 and 6.1.1.1'

_computing_data_collection               'APEX4'
_computing_cell_refinement               'SAINT'
_computing_data_reduction                'SAINT'
_computing_structure_solution            'SUPERFLIP'
_computing_structure_refinement          'Jana2006'

#=======================================================================
# 9. ATOMIC COORDINATES AND DISPLACEMENT PARAMETERS

loop_
 _atom_site_label
 _atom_site_type_symbol
 _atom_site_fract_x
 _atom_site_fract_y
 _atom_site_fract_z
 _atom_site_adp_type
 _atom_site_U_iso_or_equiv
 _atom_site_site_symmetry_multiplicity
 _atom_site_occupancy
 _atom_site_calc_flag
 _atom_site_refinement_flags
 _atom_site_disorder_assembly
 _atom_site_disorder_group
 Te1 Te 0.333333 0.666667 0.25147(2) Uani 0.00933(4) 2 1 d . . .
 Ni1 Ni 0 0 0 Uani 0.00825(6) 1 1 d . . .

loop_
 _atom_site_aniso_label
 _atom_site_aniso_type_symbol
 _atom_site_aniso_U_11
 _atom_site_aniso_U_22
 _atom_site_aniso_U_33
 _atom_site_aniso_U_12
 _atom_site_aniso_U_13
 _atom_site_aniso_U_23
 Te1 Te 0.00889(5) 0.00889(5) 0.01021(5) 0.00445(2) 0 0
 Ni1 Ni 0.00744(7) 0.00744(7) 0.00987(10) 0.00372(3) 0 0

#=======================================================================
# 10. MOLECULAR GEOMETRY

loop_
 _geom_bond_atom_site_label_1
 _geom_bond_atom_site_label_2
 _geom_bond_site_symmetry_1
 _geom_bond_site_symmetry_2
 _geom_bond_distance
 Te1 Ni1 . .     2.59019(5)
 Te1 Ni1 . 1_565 2.59019(5)
 Te1 Ni1 . 1_665 2.59019(5)

loop_
 _geom_angle_atom_site_label_1
 _geom_angle_atom_site_label_2
 _geom_angle_atom_site_label_3
 _geom_angle_site_symmetry_1
 _geom_angle_site_symmetry_2
 _geom_angle_site_symmetry_3
 _geom_angle
 Ni1 Te1 Ni1 .     . 1_565 96.186(3)
 Ni1 Te1 Ni1 .     . 1_665 96.186(3)
 Ni1 Te1 Ni1 1_565 . 1_665 96.186(3)
 Te1 Ni1 Te1 1_445 . 1_545 96.1861(19)
 Te1 Ni1 Te1 1_445 . .     96.1861(19)
 Te1 Ni1 Te1 1_445 . 4_445 83.8139(19)
 Te1 Ni1 Te1 1_445 . 4_455 83.8139(19)
 Te1 Ni1 Te1 1_445 . 4_555 180
 Te1 Ni1 Te1 1_545 . .     96.1862(19)
 Te1 Ni1 Te1 1_545 . 4_445 83.8139(19)
 Te1 Ni1 Te1 1_545 . 4_455 180
 Te1 Ni1 Te1 1_545 . 4_555 83.8139(19)
 Te1 Ni1 Te1 .     . 4_445 180
 Te1 Ni1 Te1 .     . 4_455 83.8139(19)
 Te1 Ni1 Te1 .     . 4_555 83.8139(19)
 Te1 Ni1 Te1 4_445 . 4_455 96.1861(19)
 Te1 Ni1 Te1 4_445 . 4_555 96.1861(19)
 Te1 Ni1 Te1 4_455 . 4_555 96.1862(19)

\end{Verbatim}
\newpage

6. CIF of crystal \#0 : \\

\begin{Verbatim}

#=======================================================================
data_global
#=======================================================================

_audit_creation_method                   'Jana2006 Version : 04/12/2022'

#=======================================================================
data_I
#=======================================================================

# 5. CHEMICAL DATA

_chemical_formula_sum                    'Ni Te2'
_chemical_formula_weight                 313.9
_chemical_absolute_configuration         .

#=======================================================================
# 6. CRYSTAL DATA

_symmetry_cell_setting                   trigonal
_symmetry_space_group_name_H-M           'P -3 m 1'
_symmetry_space_group_name_Hall          '-P 3 2"'
_symmetry_Int_Tables_number              164

loop_
 _space_group_symop_id
 _space_group_symop_operation_xyz
 1   x,y,z
 2   -y,x-y,z
 3   -x+y,-x,z
 4   y,x,-z
 5   x-y,-y,-z
 6   -x,-x+y,-z
 7   -x,-y,-z
 8   y,-x+y,-z
 9   x-y,x,-z
 10  -y,-x,z
 11  -x+y,y,z
 12  x,x-y,z

_cell_length_a                           3.85540(10)
_cell_length_b                           3.85540(10)
_cell_length_c                           5.2671(2)
_cell_angle_alpha                        90
_cell_angle_beta                         90
_cell_angle_gamma                        120
_cell_volume                             67.802(4)

loop_
 _twin_individual_id
 _twin_individual_mass_fraction_refined
 _twin_individual_twin_matrix_11
 _twin_individual_twin_matrix_12
 _twin_individual_twin_matrix_13
 _twin_individual_twin_matrix_21
 _twin_individual_twin_matrix_22
 _twin_individual_twin_matrix_23
 _twin_individual_twin_matrix_31
 _twin_individual_twin_matrix_32
 _twin_individual_twin_matrix_33
 1 1 1  0  0  0  1 0  0 0 1
 2 0 1 -1  0  1  0 0  0 0 1

_cell_formula_units_Z                    1

_cell_measurement_reflns_used            7625
_cell_measurement_theta_min              18.6
_cell_measurement_theta_max              72
_cell_measurement_temperature            293

_exptl_crystal_density_diffrn            7.6875
_exptl_crystal_F_000                     132
_exptl_absorpt_coefficient_mu            27.861
_exptl_crystal_description
 'plate-like with about elliptical surface, approximated as flat cylinder'
_exptl_crystal_size_max                  0.50
_exptl_crystal_size_mid                  0.50
_exptl_crystal_size_min                  0.023
_exptl_crystal_size_rad                  0.103
_exptl_crystal_colour                    'black'
_exptl_absorpt_correction_type           Gaussian
_exptl_absorpt_process_details           'Jana2006'
_exptl_absorpt_correction_T_min          0.057
_exptl_absorpt_correction_T_max          0.558

loop_
 _exptl_crystal_face_index_h
 _exptl_crystal_face_index_k
 _exptl_crystal_face_index_l
 _exptl_crystal_face_perp_dist
  0  0  1 0.0115
  0  0 -1 0.0115
  1  0  0 0.25
 -1  0  0 0.25
  0  1  0 0.25
  0 -1  0 0.25
  1  1  0 0.25
 -1 -1  0 0.25
  1 -1  0 0.25
 -1  1  0 0.25
  1  2  0 0.25
 -1 -2  0 0.25
  2  1  0 0.25
 -2 -1  0 0.25
  1 -2  0 0.25
 -1  2  0 0.25
  2 -1  0 0.25
 -2  1  0 0.25

#=======================================================================
# 7. EXPERIMENTAL DATA

_diffrn_ambient_temperature              293
_diffrn_source_voltage                   '50 kV'
_diffrn_source_current                   '30 mA'
_diffrn_radiation_type                   'Mo K\a'
_diffrn_radiation_source                 'X-ray tube'
_diffrn_radiation_wavelength             0.71075
_diffrn_radiation_monochromator          graphite
_diffrn_measurement_device               'four-circle diffractometer'
_diffrn_measurement_device_type          'Bruker D8 VENTURE'
_diffrn_detector                         'Photon III'
_diffrn_measurement_method               'phi and omega scans'

_diffrn_reflns_number                    12039
_diffrn_reflns_theta_min                 3.87
_diffrn_reflns_theta_max                 72.08
_diffrn_reflns_theta_full                66.32
_diffrn_measured_fraction_theta_max      0.95
_diffrn_measured_fraction_theta_full     0.98
_diffrn_reflns_av_R_equivalents          0.0487
_diffrn_reflns_av_unetI/netI             0.0022
_diffrn_reflns_limit_h_min               -8
_diffrn_reflns_limit_h_max               9
_diffrn_reflns_limit_k_min               -10
_diffrn_reflns_limit_k_max               10
_diffrn_reflns_limit_l_min               -13
_diffrn_reflns_limit_l_max               14
_diffrn_radiation_probe                  X-ray

#=======================================================================
# 8. REFINEMENT DATA

_refine_special_details
;
The Ni1 site occupancy was fixed at unity in the final structural model.
Anisotropic displacement parameters were refined for Ni1 and Te1.
The maximum residual electron-density peaks are located above and below
the Te position along the c axis.
;

_atom_sites_solution_primary             'charge flipping'
_atom_sites_solution_secondary           difmap
_atom_sites_solution_hydrogens           .

_reflns_number_total                     537
_reflns_number_gt                        537
_reflns_threshold_expression             'I>3\s(I)'

_refine_ls_structure_factor_coef         Fsqd
_refine_ls_R_factor_gt                   0.0252
_refine_ls_wR_factor_gt                  0.0753
_refine_ls_R_factor_all                  0.0252
_refine_ls_wR_factor_ref                 0.0753
_refine_ls_goodness_of_fit_ref           3.39
_refine_ls_goodness_of_fit_gt            3.39
_refine_ls_number_reflns                 537
_refine_ls_number_parameters             7
_refine_ls_number_restraints             0
_refine_ls_number_constraints            0
_refine_ls_weighting_scheme              sigma
_refine_ls_weighting_details             'w=1/(\s^2^(I)+0.0004I^2^)'
_refine_ls_shift/su_max                  0.0006
_refine_ls_shift/su_mean                 0.0001
_refine_diff_density_max                 7.38
_refine_diff_density_min                 -11.41
_refine_ls_extinction_method
 'B-C type 1 Lorentzian isotropic (Becker & Coppens, 1974)'
_refine_ls_extinction_coef               2500(200)

loop_
 _atom_type_symbol
 _atom_type_scat_dispersion_real
 _atom_type_scat_dispersion_imag
 _atom_type_scat_source
 Ni   0.3393  1.1124
 'International Tables Vol C tables 4.2.6.8 and 6.1.1.1'
 Te  -0.5308  1.6751
 'International Tables Vol C tables 4.2.6.8 and 6.1.1.1'

_computing_data_collection               'APEX4'
_computing_cell_refinement               'SAINT'
_computing_data_reduction                'SAINT'
_computing_structure_solution            'SUPERFLIP'
_computing_structure_refinement          'Jana2006'

#=======================================================================
# 9. ATOMIC COORDINATES AND DISPLACEMENT PARAMETERS

loop_
 _atom_site_label
 _atom_site_type_symbol
 _atom_site_fract_x
 _atom_site_fract_y
 _atom_site_fract_z
 _atom_site_adp_type
 _atom_site_U_iso_or_equiv
 _atom_site_site_symmetry_multiplicity
 _atom_site_occupancy
 _atom_site_calc_flag
 _atom_site_refinement_flags
 _atom_site_disorder_assembly
 _atom_site_disorder_group
 Te1 Te 0.333333 0.666667 0.25140(2) Uani 0.00880(4) 2 1 d . . .
 Ni1 Ni 0 0 0 Uani 0.00772(7) 1 1 d . . .

loop_
 _atom_site_aniso_label
 _atom_site_aniso_type_symbol
 _atom_site_aniso_U_11
 _atom_site_aniso_U_22
 _atom_site_aniso_U_33
 _atom_site_aniso_U_12
 _atom_site_aniso_U_13
 _atom_site_aniso_U_23
 Te1 Te 0.00801(6) 0.00801(6) 0.01040(7) 0.00400(3) 0 0
 Ni1 Ni 0.00662(8) 0.00662(8) 0.00993(11) 0.00331(4) 0 0

#=======================================================================
# 10. MOLECULAR GEOMETRY

loop_
 _geom_bond_atom_site_label_1
 _geom_bond_atom_site_label_2
 _geom_bond_site_symmetry_1
 _geom_bond_site_symmetry_2
 _geom_bond_distance
 Te1 Ni1 . .     2.59000(6)
 Te1 Ni1 . 1_565 2.59000(6)
 Te1 Ni1 . 1_665 2.59000(6)

loop_
 _geom_angle_atom_site_label_1
 _geom_angle_atom_site_label_2
 _geom_angle_atom_site_label_3
 _geom_angle_site_symmetry_1
 _geom_angle_site_symmetry_2
 _geom_angle_site_symmetry_3
 _geom_angle
 Ni1 Te1 Ni1 .     . 1_565 96.195(3)
 Ni1 Te1 Ni1 .     . 1_665 96.195(3)
 Ni1 Te1 Ni1 1_565 . 1_665 96.195(3)
 Te1 Ni1 Te1 1_445 . 1_545 96.195(2)
 Te1 Ni1 Te1 1_445 . .     96.195(2)
 Te1 Ni1 Te1 1_445 . 4_445 83.805(2)
 Te1 Ni1 Te1 1_445 . 4_455 83.805(2)
 Te1 Ni1 Te1 1_445 . 4_555 180
 Te1 Ni1 Te1 1_545 . .     96.195(2)
 Te1 Ni1 Te1 1_545 . 4_445 83.805(2)
 Te1 Ni1 Te1 1_545 . 4_455 180
 Te1 Ni1 Te1 1_545 . 4_555 83.805(2)
 Te1 Ni1 Te1 .     . 4_445 180
 Te1 Ni1 Te1 .     . 4_455 83.805(2)
 Te1 Ni1 Te1 .     . 4_555 83.805(2)
 Te1 Ni1 Te1 4_445 . 4_455 96.195(2)
 Te1 Ni1 Te1 4_445 . 4_555 96.195(2)
 Te1 Ni1 Te1 4_455 . 4_555 96.195(2)




# start Validation Reply Form
_vrf_DIFMN02_I
;
PROBLEM: The minimum difference density is < -0.1*ZMAX*2.00
RESPONSE: The measurement was performed on a relatively large crystal containing
          the heavy element Te. Consequently, strong absorption effects were
          expected. Despite application of a Gaussian integration absorption
          correction, residual absorption-related systematic errors could not
          be fully eliminated, which likely contribute to the pronounced
          negative residual electron density.
;
_vrf_DIFMN03_I
;
PROBLEM: The minimum difference density is < -0.1*ZMAX*0.75
RESPONSE:  See the response to the related alert concerning the minimum
          difference density. The relatively large Te containing crystal
          exhibited strong absorption effects. Residual systematic errors
          remaining after the Gaussian integration absorption correction
          likely account for the pronounced negative difference density.
;
_vrf_DIFMX02_I
;
PROBLEM: The maximum difference density is > 0.1*ZMAX*0.75
RESPONSE:  The relatively large Te containing crystal exhibited strong
          absorption effects. Residual absorption-related systematic errors
          remaining after the Gaussian integration absorption correction
          likely contribute to the pronounced positive difference density.
;
_vrf_GOODF01_I
;
PROBLEM: The least squares goodness of fit parameter lies
RESPONSE: The elevated goodness-of-fit value is attributed to residual
          systematic errors, primarily from the strong absorption effects
          associated with the relatively large Te-containing crystal.
          To avoid detector saturation caused by very intense reflections,
          all data were collected using an attenuated incident beam and
          correspondingly longer counting times. Despite these limitations,
          the structural model refined satisfactorily.
;
_vrf_PLAT098_I
;
PROBLEM: Large Reported Min. (Negative) Residual Density -11.41 eA-3
RESPONSE: The large negative residual density is attributed to residual
          absorption-related systematic errors. The relatively large crystal
          contains the heavy element Te and consequently exhibited strong
          absorption effects. These effects could not be fully eliminated by
          the Gaussian integration absorption correction.
;
_vrf_PLAT097_I
;
PROBLEM: Large Reported Max. (Positive) Residual Density 7.38 eA-3
RESPONSE: The large positive residual density is attributed to residual
          absorption-related systematic errors. The relatively large crystal
          contains the heavy element Te and consequently exhibited strong
          absorption effects. These effects could not be fully eliminated by
          the Gaussian integration absorption correction.
;
# end Validation Reply Form
\end{Verbatim}

\end{document}